\documentclass[letterpaper, 10 pt, conference]{ieeeconf}  

\IEEEoverridecommandlockouts                              

\usepackage{amsmath} 
\usepackage{amssymb}  
\usepackage{xcolor}
\usepackage{comment}
\usepackage{todonotes}
\usepackage{epstopdf}
\usepackage{subcaption}

\usepackage[linesnumbered,lined,ruled]{algorithm2e}

\makeatletter
\providecommand*{\diff}%
{\@ifnextchar^{\DIfF}{\DIfF^{}}}
\def\DIfF^#1{%
\mathop{\mathrm{\mathstrut d}}%
\nolimits^{#1}\gobblespace}
\def\gobblespace{%
\futurelet\diffarg\opspace}
\def\opspace{%
\let\DiffSpace\!%
\ifx\diffarg(%
\let\DiffSpace\relax
\else
\ifx\diffarg[%
\let\DiffSpace\relax
\else
\ifx\diffarg\{%
\let\DiffSpace\relax
\fi\fi\fi\DiffSpace}

\title{\LARGE \bf
Online Identification of Stochastic Continuous-Time\\ Wiener Models Using Sampled Data
}

\author{Mohamed Abdalmoaty$^{1}$, Efe C. Balta$^{2}$, John Lygeros$^{1}$, and Roy S. Smith$^{1}$
\thanks{This work has been supported by the Swiss National Science Foundation under NCCR Automation (grant agreement  $51\text{NF}40\_180545$)}
\thanks{$^{1}$Automatic Control Laboratory (IfA), Swiss Federal Institute of Technology (ETH Z\"{u}rich), 8092 Z\"{u}rich, Switzerland,
        {\tt\footnotesize \{mabdalmoaty,jlygeros,rsmith\}@control.ee.ethz.ch}\newline%
        $^{2}$Control and Automation Group, Inspire AG, 8005 Z\"{u}rich, Switzerland {\tt\footnotesize efe.balta@inspire.ch}}
}

\begin{document}
\maketitle
\thispagestyle{empty}
\pagestyle{empty}

\begin{abstract}
It is well known that ignoring the presence of stochastic disturbances in the identification of stochastic Wiener models leads to asymptotically biased estimators. On the other hand, optimal statistical identification, via likelihood-based methods, is sensitive to the assumptions on the data distribution and is usually based on relatively complex sequential Monte Carlo algorithms. We develop a simple recursive online estimation algorithm based on an output-error predictor, for the identification of continuous-time stochastic parametric Wiener models through stochastic approximation. The method is applicable to generic model parameterizations and, as demonstrated in the numerical simulation examples, it is robust with respect to the assumptions on the spectrum of the disturbance process. 
\end{abstract}

\section{INTRODUCTION}\label{sec:intro}
Online system identification is a classical problem in the Systems and Control literature. Several methods and algorithms were developed in parallel to the development of adaptive control techniques; see \cite{Hammond1965}, \cite{Eykhoff1974}.
Apart from its direct practical interest, online identification has close connections to nonlinear filtering, stochastic approximation, and learning and adaptation;  see \cite{young2011,Goodwin2014,chen2014}, and \cite{Ljung1983} for a unified framework based on prediction errors.

The majority of classical prediction error methods (PEM) were designed for linear stochastic models, or nonlinear deterministic models, permitting disturbances or noise solely at the output, \cite[Ch. 11]{Ljung1999}, \cite[Ch. 9]{Soederstroem1989}, \cite[Ch. 8]{Goodwin2014}. For general nonlinear stochastic discrete-time state-space models, likelihood-based estimators such as the maximum-likelihood estimator or the maximum-a-posteriori estimator are usually used. Both offline and online implementations of such estimators rely on (particle) sequential Monte Carlo approximations, see for example \cite{Olsson2020}, the survey \cite{kantas2015}, and the references therein. 
Similar algorithms for continuous-time nonlinear time-series models were proposed as offline methods using sampled measurements \cite{Singer2018} 
and as online algorithms using continuous-time observations \cite{Beskos2021,Surace2018}. Beyond computational constraints such as particle degeneracy, likelihood-based techniques are sensitive to model misspecification. The maximum-likelihood estimator deviates from its optimal asymptotic properties when there are discrepancies in the data distribution \cite{Abdalmoaty2020b}.

Most of the existing identification methods have only asymptotic approximation guarantees.
This is mainly due to the intractability of the finite-sample distributions of the estimators, which are usually highly nonlinear in the data. Yet, asymptotic theory is relatively easy to develop under general regularity conditions. Some major advantages of asymptotic analysis are the applicability to fairly general model parameterizations in both continuous- and discrete-time, and the ability to work with model misspecification; a clear drawback is the lack of finite-sample guarantees.

System identification of linear systems from a single input-output trajectory using linear least-squares estimators with non-asymptotic guarantees has been studied in~\cite{oymak2019non} and \cite{simchowitz2020improper}. Most of the existing work employs persistently exciting stochastic input sequences from a predetermined distribution to provide convergence guarantees; see~\cite{matni2019tutorial} for an overview of recent results.  
Nonlinear closed-loop policies with excitation noise for the identification of discrete-time linear dynamics are proposed in~\cite{li2023non}.
Additionally, generalized linear, nonlinear~\cite{foster2020learning}, and piecewise-affine systems~\cite{mania2020active} have been studied. However, existing algorithms often focus on discrete-time dynamics and are not easily adaptable to continuous-time. Furthermore, while the results may be applicable, recursive online identification is often neglected.

In this work, we focus on the online estimation problem of continuous-time stochastic parametric Wiener models using noisy samples of the output signal. A Wiener model represents a specific case within the category of block-oriented models, comprising a linear dynamical model, followed by a static nonlinearity at its output. It has found application in different scientific and engineering domains \cite{Bai2010book} as it can approximate a large class of nonlinear systems \cite{Boyd1985}. Because the dynamic component is linear, exact time-discretization is possible when the inter-sample behavior of the input signal is known. Moreover, the stability of a Wiener model is dictated by the stability of the linear component. Previous work on recursive online identification of Wiener models, as in \cite{Wigren1993} and  \cite{Wigren1994}, has only considered the deterministic case in discrete time. In \cite{Wigren2023}, an online PEM algorithm for deterministic nonlinear continuous-time state-space models was explored using an Euler approximate discretization scheme.

We propose a relatively simple online parameter estimation algorithm for the class of continuous-time stochastic Wiener models. We utilize an output-error predictor and adopt an input-output approach, accommodating a sampled data scenario with additive output measurement noise. 
 The algorithm is developed using an online prediction error framework, which is furnished with a powerful asymptotic theory. Through numerical simulation examples, we showcase the algorithm's performance, especially in cases where the disturbance model is incorrectly specified.
 It should be noted that even though we consider continuous-time models, the approach is directly applicable to discrete-time models.

The outline of the paper is as follows. The problem formulation and assumptions are given in Section~\ref{sec:problem}. The proposed approach and the overall algorithm are described in Section~\ref{sec:approach} together with some theoretical motivation. 
Section~\ref{sec:examples} provides numerical simulations. Lastly,  conclusions and future work are given in Section~\ref{sec:conclusions}.

\section{Problem Formulation}\label{sec:problem}
We consider a class of stochastic parametric time-invariant continuous-time Wiener models
\begin{equation}\label{eq:model}
\begin{aligned}
    \diff w(t) & = A(\theta) w(t) dt + B(\theta) \diff \beta(t),\\
    x(t) & = G(p;\theta) u(t) + C(\theta) w(t),\\
    y(t) & = f(x(t);\theta),
\end{aligned}
\end{equation}
where $G(p;\theta)$ is a continuous-time transfer operator, $p$ is the differential operator, $f(\cdot;\theta)$ is a static parametric function, $\theta \in \Theta \subset \mathbb{R}^d$ is a parameter vector, $u(t)$ is the input signal, $y(t)$ is the output signal, $x(t)$ is a latent signal, and $w(t) \in \mathbb{R}^{n_w}$ is a disturbance driven by a Wiener process $\beta(t)$. Additionally, $A(\theta), B(\theta), C(\theta)$ are parametric matrices of appropriate sizes.
We assume the output is measured at discrete-time instances $t_k$ with additive measurement noise,
\begin{equation}
    y_k = y(t_k) + v_k, \quad k = 1, 2, 3, \dots
\end{equation}
Without loss of generality, we assume that $v_k$ has a zero-mean value for all $k$ (non-zero constant mean values can be included in the parameterization of $f$).   
We also assume that the input signal is known exactly as a continuous-time signal, and therefore the data set available at time $t_N$ is
\[
D_N : = \big\{\left(y_k, u(t)\right) : k= 1, \dots, N, \; t \in [t_1,t_N ]  \big\}.
\]
Moreover, the data is collected open-loop so that $u$ is independent of $w$ and $v$. To ensure that the data collection process is well-posed,  
 model \eqref{eq:model} is assumed to be stable for all $\theta\in\Theta$.
To simplify the exposition, we confine our discussion to the single-input single-output scenario using a constant sampling period $\Delta$. 
However, the suggested method is applicable to the more general case with multiple inputs and outputs with irregular sampling times.
The choice of a transfer operator parametrization is
\begin{equation}
    G(p;\theta) \!=\! \frac{\sum_{j=0}^m c_j p^j}{p^n+\sum_{j=0}^{n-1} d_j p^j}, 
\end{equation}
where $m\leq n$,  $\theta_G \!:=\! [ c_0 \, \dots \, c_m \;d_0 \,\dots d_{n-1} ]^\top \!\in\! \mathbb{R}^{d_G}$ and $d_G = n+m+1$. 
Other parametrizations, e.g., state space (canonical or not), are also possible.

The parametrization of the It\^{o} stochastic differential equation used to model the disturbance  $w$ is done separately from  $G$ and  $f$, allowing for the possibility of misspecification only in the disturbance model. The matrices $A(\theta)$, $B(\theta)$, and $C(\theta)$ assume a state-space parametrization that should be identifiable from the marginal second-order moments of $y$. This typically means that only a few parameters in one of the matrices can be estimated.\footnote{In general, estimating a fully parameterized black-box disturbance model requires a different cost function than the one adopted in this work.} Nevertheless, the proposed approach can naturally handle cases where $G$, $f$ and the model of $w$ are jointly parameterized (see Section \ref{ex1}). We order the entries of the parameter vector as $\theta = [\theta_G^\top\; \theta_w^\top \; \theta_f^\top]^\top$, where $\theta_w \in \mathbb{R}^{d_w}$ are parameters appearing in the disturbance model, and $\theta_f \in \mathbb{R}^{d_g}$ are  parameters of $f$.

The main objective of the paper is the construction of an online estimation algorithm for $\theta$ that, based on the knowledge of the inter-sample behavior of $u$, maps the current measurement $y_k$ to an estimate $\hat{\theta}_k$. Special emphasis is placed on the asymptotic properties of the resulting sequence of estimators $\{\hat{\theta}_k\}_{k\geq1}$. When the system is time-invariant, an appropriate algorithm design ensures the almost sure convergence to a subset of $\Theta$. 

\section{Proposed Approach}\label{sec:approach}

\subsection{Offline method: The OE-QPEM estimator}
The Output-Error Quadratic PEM (OE-QPEM)\footnote{OE because the predictor does not depend on previous output measurements $y_k$, QPEM because $V_N$ is defined using squared Euclidean norm.} estimator based on  $D_N$ is 
 defined as the minimizer of 
\begin{equation}\label{eq:cost_function}
V_N(\theta) := \frac{1}{N} \sum_{k=1}^N \frac{1}{2} \left( y_k - \mathbb{E}\left[y_k|(u(s))_{s=t_1}^{t_k};\theta\right]\right)^2,
\end{equation}
over a suitable compact subset $\Theta \subset \mathbb{R}^d$. 
The expectation operator is with respect to the process disturbance $w$ and the measurement noise $v$, and is conditioned on the known input signal $u$. This estimator was proposed in~\cite{Abdalmoaty2019} with the analysis of its convergence and asymptotic properties. The estimator provides a computationally simpler alternative to likelihood-based methods, as it does not require the computation (or approximation) of the predictive densities $p(y_k | y_1, \dots, y_{k-1} , (u(s))_{s=t_1}^{t_{k}}; \theta)$ of the model's output. Yet, the decrease in statistical efficiency is typically modest, and is often outweighed by the computational simplicity it offers (as shown e.g. in~\cite{Abdalmoaty2018b} and \cite{Abdalmoaty2019}), as well as the applicability to complex models; see e.g., \cite{bereza2022stochastic} where the offline OE-QPEM estimator is applied to stochastic DAE models.

\subsection{Proposed identification method}\label{sec:algorithm}
We propose an online algorithm that yields estimators with the same asymptotic properties as the offline OE-QPEM estimator. In a similar vein to \eqref{eq:cost_function}, we seek to minimize
\begin{equation}\label{eq:asymptotic_cost}
    V(\theta) = \frac{1}{2} \mathbb{E}[\varepsilon_k^2(\theta)],
\end{equation}
where  $\varepsilon_k(\theta) :=  y_k - \mathbb{E}\left[y_k|(u(s))_{s=t_1}^{t_k};\theta\right]$ is the prediction error, using stochastic approximation. 
Let us denote the predictor and the gradient vector of the prediction error with respect to $\theta$ as \vspace{-0.2cm}
\[
\begin{aligned}
\hat{y}_k (\theta)& = \mathbb{E}\left[y_k|(u(s))_{s=t_1}^{t_k};\theta\right],\\
\psi_k(\theta) &= \frac{\diff }{\diff \theta} \varepsilon_k(\theta) = -\frac{\diff }{\diff \theta} \hat{y}_k(\theta),
\end{aligned}
\]
respectively. 
Then 
\[
	\begin{aligned}
		V'(\theta) & = \frac{\diff }{\diff \theta} \frac{1}{2} \mathbb{E}[\varepsilon_k^2(\theta)] =  \mathbb{E}\left[ \psi_k(\theta) \left(y_k - \hat{y}_k(\theta)\right) \right], 
	\end{aligned}
\]
where we have allowed the interchange of expectation and differentiation. Notice that the outer expectation pertains to the underlying probability space of the data (unknown), while the inner expectations, defining $\hat{y}_k(\theta)$ and $\psi_k(\theta)$, are with respect the Wiener process $\beta(t)$ in \eqref{eq:model}. Then, minimizing $V(\theta)$ can be achieved by solving the system of equations
\begin{equation}\label{eq:estimating_function}
    \mathbb{E}\left[ \psi_k(\theta)(y_k - \mathbb{E}\left[y_k|(u(s))_{s=t_1}^{t_k};\theta\right]\right] = 0.
\end{equation}
Applying the Robbins-Monro stochastic approximation scheme \cite{Robbins1951}, we obtain the following stochastic gradient descent recursion
\begin{equation}\label{eq:sgd}
\hat{\theta}_k = \hat{\theta}_{k-1} + \gamma_k \psi_k(\hat{\theta}_{k-1})\left(y_k - \hat{y}_k(\hat{\theta}_{k-1})\right),
\end{equation}
 in which $\gamma_k$ are positive scalars tending to zero sufficiently slowly as $k$ grows.
There are two primary challenges in computing the OE predictor $\hat{y}_k(\hat{\theta}_{k-1})$ and the gradient vector $\psi_k(\hat{\theta}_{k-1})$. The first is the evaluation of the expected values with respect to $\beta$, and the second is doing so online in a recursive manner. The solution to the first challenge is generic in nature, while the second naturally depends on the choice of model class and parameterization. 

The main idea of our approach is to compute the OE predictor $\hat{y}_k(\hat{\theta}_{k-1})$ and the corresponding prediction error gradient vector $\psi_k(\hat{\theta}_{k-1})$ in \eqref{eq:sgd}  by simulating \eqref{eq:model} and its output gradient filters using two independent Wiener processes $\beta^{(y)}(t)$ and $\beta^{(\psi)}(t)$, respectively, at $\theta = \hat{\theta}_{k-1}$.
A variety of other methods can be employed to compute {online} predictors for \eqref{eq:model}, but our approach provides a simple implementation with favorable properties as discussed below.

The outputs of these two simulations, denoted $y_{1,k}(\hat{\theta}_{k-1})$ and $\psi_{1,k}(\hat{\theta}_{k-1})$, are unbiased estimators of the OE predictor and the corresponding prediction error gradient vector at time $t_k$, evaluated at $\hat{\theta}_{k-1}$. Moreover, they are independent by construction. This important property means that the  vector
\begin{equation}\label{eq:unbiased_estimator-of_the_EF}
\begin{aligned}
\psi_{1,k}(\hat{\theta}_{k-1}) \left(y_k - y_{1,k}(\hat{\theta}_{k-1})\right)
\end{aligned}  
\end{equation}
is an unbiased estimator of the estimating function in \eqref{eq:estimating_function}. 
While  only two Wiener processes are needed,  the performance of the algorithm may be improved by considering the average of $M \geq 1$ independent simulations:
\[
\begin{aligned}
    \bar{y}_k(\hat{\theta}_{k-1}) &= \frac{1}{M} \sum_{m=1}^M  y_{m,k}(\hat{\theta}_{k-1}),\\
    \bar{\psi}_k(\hat{\theta}_{k-1}) &= \frac{1}{M} \sum_{m=1}^M  \psi_{m,k}(\hat{\theta}_{k-1}).
\end{aligned}
\]
These unbiased estimators can be thought of as ``measurements" of $\hat{y}_k(\hat{\theta}_{k-1})$ and $\psi_k(\hat{\theta}_{k-1})$, respectively. With this in mind, the Robbins-Monro scheme applied to \eqref{eq:model} gives
\[
\hat{\theta}_k = \hat{\theta}_{k-1} + \gamma_k \bar{\psi}_k(\hat{\theta}_{k-1})\left(y_k - \bar{y}_k(\hat{\theta}_{k-1})\right).
\]
The number of simulations $M$ could be thought of as a tuning parameter of the algorithm, just like the sequence of gains $\gamma_k$. Note that  $M$ can be either fixed or changed with $k$.

To further improve the convergence properties, a stochastic Newton direction can be used. The Hessian of $V(\theta)$ is 
\[
V''(\theta) = \frac{\diff^2}{\diff \theta^2} \frac{1}{2} \mathbb{E}[\varepsilon_k^2(\theta)] = \mathbb{E}[\psi_k(\theta) \psi^\top_k(\theta)].
\]
 For a fixed $\theta$, the Hessian can be determined as the solution $R$ of the system of equations
$\mathbb{E}[\psi_k(\theta) \psi_k^\top(\theta) - R ]  = 0$ 
where the expectation is with respect to the data distribution. Using the ideas outlined above, we arrive at the following stochastic Newton algorithm
\[
\begin{aligned}
R_k &= R_{k-1}+ \gamma_k \left[\bar{\psi}_k(\theta_{k-1})\left[\bar{\psi}_k(\theta_{k-1})\right]^\top - R_{k-1}\right],\\
\hat{\theta}_k &= \left[\hat{\theta}_{k-1} + \gamma_k R^{-1}_k \bar{\psi}_k(\hat{\theta}_{k-1})\left(y_k -\bar{y}_k(\hat{\theta}_{k-1})\right)\right]_\Theta,
\end{aligned}
\]
where $\left[\;\cdot\;\right]_\Theta$ is a projection operator. A particularly simple implementation defines (see \cite[(11.50)]{Ljung1999})
\[
\left[\;\theta\;\right]_\Theta := \begin{cases}
\theta \qquad &\text{if } \theta \in \Theta\\
\hat{\theta}_{k-1} &\text{if } \theta \notin \Theta.
\end{cases}
\]
The only issue with this algorithm is that it is not recursive. The estimates of the OE predictor and its gradient vector at time $t_k$ are the outputs of filters with infinite impulse responses, and hence they rely on all past data in general. This issue is fixed below using approximation processes similar to those used in classical online PEM algorithms.

\subsection{Recursive computation of the OE predictor}
We now construct a natural recursive approximation $\bar{y}_k$ of $\bar{y}_k(\hat{\theta}_{k-1})$. Define $z(t;\theta) = G(p;\theta) u(t)$, $z_k(\theta) = z(t_k;\theta)$. For the sake of clarity, we let the input be constant over the sampling interval: $u(t) = u_k$, for $k\Delta \leq t < (k+1)\Delta$.
Then the sampled-data transfer function is\footnote{This is simply zero-order hold (ZOH) sampling. The symbol  $\mathcal{L}^{-1}\{\cdot\}$ denotes the inverse Laplace transform,  $\mathcal{Z}\{\cdot\}$ denotes the $\mathrm{Z}$-transform, and $\mathrm{z}$ is the $\mathrm{Z}$-transform complex variable.}
\[
\begin{aligned}
G_\Delta(\mathrm{z}^{-1};\theta) &:= (1-\mathrm{z}^{-1}) \mathcal{Z}\left\{ \mathcal{L}^{-1}\left\{ \frac{G(p;\theta)}{s} \right\}_{t=k\Delta} \right\} \\
&=:\frac{\sum_{r=1}^{n}b_r(\theta) \mathrm{z}^{-r}}{1+\sum_{r=1}^{n}a_r(\theta) \mathrm{z}^{-r}}
\end{aligned}
\]
When the input is not constant over the sampling interval, and/or the sampling times are irregular, the model is discretized exactly by using the knowledge of the inter-sample behaviour (see Section \ref{ex1}).  
For any fixed $\theta$, the following recursion holds exactly
\[
\begin{aligned}
z_k(\theta)  = &-a_1(\theta) z_{k-1}(\theta) - \dots - a_n(\theta) z_{k-n}(\theta)\\
&+ b_1(\theta) u_{k-1} +\dots + b_n(\theta) u_{k-n}.
\end{aligned}
\]
A recursive approximation $z_k$ of $z_k(\hat{\theta}_{k-1})$ is then obtained with the current estimate $\hat{\theta}_{k-1}$, using previous values of $z_k$ as initial values. We denote the approximation of $z_k(\hat{\theta}_{k-1})$ compactly by
\begin{equation}\label{eq:approx_Gu}
z_k  =  \varphi_{k-1}^\top \eta(\hat{\theta}_{k-1}),
\end{equation}
where $\varphi_{k-1} = \begin{bmatrix} -z_{k-1}\, \dots\, -z_{k-n}\; u_k\, \dots\,  u_{k-n}\end{bmatrix}^\top$\!, and
\[
\eta(\hat{\theta}_{k-1}) = \begin{bmatrix} a_1(\hat{\theta}_{k-1}) \,\! \hdots \!\, a_n(\hat{\theta}_{k-1}) \;
                        b_1(\hat{\theta}_{k-1}) \,\! \hdots \,\! b_n(\hat{\theta}_{k-1}) \end{bmatrix}^\top\!\!.
\]
Likewise, a recursive approximation, denoted as $w_{m,k}$, is derived for $w_m(t_k; \hat{\theta}_{k-1})$. Because the model of $w$ is linear, it can be sampled exactly (in the sense that the statistical properties are identical at the sampling times; see e.g., \cite[Ch.3, Sec.10, pages 82-83]{Astroem1970}) to get
\begin{equation}
\begin{aligned}
w_{m,k+1}(\theta) &= A_\Delta(\theta)w_{m,k}(\theta) + B_\Delta(\theta) \beta_{m,k}^{(y)},\\
\beta_{m,k}^{(y)} &\overset{\text{iid}}{\sim} \mathcal{N}(0, I_{n_w})
\end{aligned}
\end{equation}
where $B_\Delta(\theta)$ is a square root of the covariance matrix given by the integral $\int_0^\Delta e^{A(\theta)s}   B(\theta) B^\top(\theta)  e^{A^\top(\theta) s} \diff s$ and $A_\Delta(\theta) = e^{A(\theta)\Delta}$. The approximation process is then computed recursively as
\begin{equation}\label{eq:approx_w_y}
w_{m,k} = A_\Delta(\hat{\theta}_{k-1}) w_{m,k-1} + B_\Delta(\hat{\theta}_{k-1}) \beta_{m,k}^{(y)}
\end{equation}
where at time $t_k$ we only need to store $\{w_{m,k}\}_{m=1}^M$. The random variables $\beta_{m,k}^{(y)}$ are sampled in run-time and not stored. 
Finally, we define
\[
    \bar{y}_k = \frac{1}{M} \sum_{m=1}^M f\left(  z_k + C(\hat{\theta}_{k-1}) w_{m,k}\,; \hat{\theta}_{k-1}\right)
\]
which only requires storing $\varphi_k$, $\{w_{m,k}\}_{m=1}^M$, and $\hat{\theta}_{k}$ at each time step.

\subsection{Recursive computation of the gradient vector}
Define $x_m(t;\theta) = z(t;\theta) + C(\theta) w_m(t;\theta)$, 
where $w_m(t;\theta)$ is driven by $\beta^{(\psi)}_m$. 
Applying the chain rule,
\begin{equation}\label{eq:grad_x}
\begin{aligned}
\partial_{\theta_j}[x_m(t;\theta)] = \partial_{\theta_j} [z(t;\theta)] &+ C_j(\theta) \, w_m(t;\theta) \\
&+ C(\theta) \partial_{\theta_j}[w_m(t;\theta)],
\end{aligned}
\end{equation}
where $C_j(\theta)$ is the entry-wise derivative of $C(\theta)$ with respect to $\theta_j$. Similarly, we have
\begin{equation}\label{eq:grad_y}
\begin{aligned}
    \partial_{\theta_j}\left[ y_m(t; \theta)\right] &= \partial_{\theta_j}\left[ f( a; \theta)\right] \big|_{a =  x_m(t;\theta)}\\
    &+  f(x_m(t; \theta),\theta) \,\partial_{\theta_j}\left[x_m(t;\theta)\right],
\end{aligned}
\end{equation}
and 
\[
\begin{aligned}
    \partial_{\theta_j} [z(t;\theta)] &= G'_j(p;\theta) u(t), \hspace{2.35cm} 1\leq j\leq d_G\\
    & =  \begin{cases}  p^j G(p;\theta) u(t),    & 1 \leq j \leq m \\
                                                         \frac{-p^j}{p^n+\sum_{j=1}^n d_j p^j} G(p;\theta) u(t), &  m+1 \leq j \leq d_G
                                          \end{cases}
\end{aligned}
\]
The gradient filters $G'_j(p;\theta)$ can be discretized similarly to $G(p;\theta)$, based on the inter-sample behaviour of $u(t)$, to get
\[
\begin{aligned}
    G'_{j}(\mathrm{z}^{-1};\theta) &:=\frac{\sum_{r=1}^{n_j}b_r^{(j)}(\theta) \mathrm{z}^{-r}}{1+\sum_{r=1}^{n_j}a^{(j)}_r(\theta) \mathrm{z}^{-r}},
\end{aligned}
\]
in which $n_j = n$ for $1\leq j \leq m$, while $n_j = 2n$ for $m+1 \leq j \leq d_G$. 

Analogous to \eqref{eq:approx_Gu}, a recursive approximation $z_k^{(j)}$ of $\partial_{\theta_j} [z(t_k;\theta)]$ is defined as
\[
z_k^{(j)} = [\varphi_{k-1}^{(j)}]^\top \eta_{j}(\hat{\theta}_{k-1}),
\]
in which $\varphi_{k-1}^{(j)} =\begin{bmatrix} -z_{k-1}^{(j)} \, \dots\, -z_{k-n_j}^{(j)} \; u_{k-1}\, \dots\,  u_{k-n_j}\end{bmatrix}^\top$,  
\begingroup\makeatletter\def\f@size{9}\check@mathfonts
\[
    \eta_{j}(\hat{\theta}_{k-1}) = \begin{bmatrix} a_1^{(j)}(\hat{\theta}_{k-1}) \,\! \hdots \!\, a_{n_j}^{(j)}(\hat{\theta}_{k-1}) \;
                        b_1^{(j)}(\hat{\theta}_{k-1}) \,\! \hdots \!\, b_{n_j}^{(j)}(\hat{\theta}_{k-1}) \end{bmatrix}^\top\!\!.
\]
\endgroup

On the other hand, the gradients $\partial_{\theta_j}[w_m(t;\theta)]$, denoted as $w_{m}^{(j)}(t;\theta)$ in the sequel, are obtained by differentiating the stochastic integral equations defining $w_m(t;\theta)$ with respect to $\theta$. It can be shown that they satisfy
\begin{equation}\label{eq:grad_filters_w}
\begin{aligned}
    \diff \zeta^{(j)}(t;\theta) = F^{(j)}(\theta) \zeta^{(j)}(t;\theta) \diff t + L^{(j)}(\theta) \diff \beta^{(\psi)}_m(t)
\end{aligned}
\end{equation}
in which $\zeta^{(j)}(t;\theta) = \begin{bmatrix} [w_m(t;\theta)]^\top &[w^{(j)}_m(t;\theta)]^\top\end{bmatrix}^\top$,  with the following drift and dispersion matrices
\[
    F^{(j)}(\theta) =  \begin{bmatrix}
        A(\theta) & 0\\
        A_j(\theta) & A(\theta)
    \end{bmatrix}, \qquad 
    L^{(j)}(\theta) =  \begin{bmatrix}B(\theta) \\ B_j(\theta) \end{bmatrix}. 
\]
Here, $A_j(\theta)$ and $B_j(\theta)$ are defined similarly to $C_j(\theta)$. Notice that $\beta^{(\psi)}_m$ in~\eqref{eq:grad_filters_w}
and $\beta^{(y)}_m$ in \eqref{eq:approx_w_y} are two independent Wiener processes by definition. The sampled versions  of $\zeta^{(j)}(t;\theta)$ are 
 \[
 \begin{aligned}
 \zeta_{m,k+1}^{(j)}(\theta) &= F^{(j)}_\Delta(\theta) \zeta_{m,k}^{(j)}(\theta) + L^{(j)}_\Delta(\theta) \beta_{m,k}^{(\psi)}\\
\beta_{m,k}^{(\psi)} &\overset{\text{iid}}{\sim} \mathcal{N}(0, I_{2n_w}),
 \end{aligned}
 \]
 where  $F^{(j)}_\Delta(\theta) = e^{F^{(j)}(\theta)\Delta}$, and $L^{(j)}_\Delta(\theta)$ is a square root of the covariance matrix given by the integral $\int_0^\Delta e^{F^{(j)}(\theta)s}   L^{(j)}(\theta) [L^{(j)}(\theta)]^\top  e^{[{F^{(j)}(\theta)]^\top} s} \diff s$. For irregular sampling times, $\Delta$ is simply replaces by $\Delta_k= t_{k+1} - t_k$. 
 
 The approximation process is then computed as
 \[
 \begin{aligned}
  \zeta_{m,k+1}^{(j)} &= F^{(j)}_\Delta(\hat{\theta}_{k-1}) \zeta_{m,k}^{(j)} + L^{(j)}_\Delta(\hat{\theta}_{k-1}) \beta_{m,k}^{(\psi)},\\
 {w}_{m,k} &= [\zeta_{m,k}^{(j)}]_{1:n_w}, \qquad {w}_{m,k}^{(j)} = [\zeta_{m,k}^{(j)}]_{n_w:2n_w}.
 \end{aligned}
 \]
Notice that, (i) despite using the same symbol as in \eqref{eq:approx_w_y}, ${w}_{m,k}$ here is driven by $\beta^{(\psi)}_{m,k}$ (which are sampled in run-time) and is independent of that in \eqref{eq:approx_w_y}, and (ii) ${w}_{m,k}$ is the same for all $j$.

The recursive approximations of \eqref{eq:grad_x} and \eqref{eq:grad_y} are
\[
\begin{aligned}
    x_{m,k}^{(j)} &= z_k^{(j)}+ C_j(\hat{\theta}_{k-1}) w_{m,k} + C(\hat{\theta}_{k-1}) w_{m,k}^{(j)},\\
    y^{(j)}_{m,k} &= \partial_{\theta_j}\left[ f( a; \theta)\right] \Bigr|_{\substack{a =  x_{m,k}\\ \theta= \hat{\theta}_{k-1}}}+ f(x_{m,k};\hat{\theta}_{k-1}) x_{m,k}^{(j)}.
\end{aligned}
\]
Finally, with $\psi_{m,k} = -\begin{bmatrix} y^{(1)}_{m,k} \!&\! y^{(2)}_{m,k}\! &\! \dots \!&\! y^{(d)}_{m,k} \end{bmatrix}^\top$, 
\[
    \bar{\psi}_k = \frac{1}{M}  \sum_{m=1}^M \psi_{m,k}
\]
which provides a recursive approximation of $\bar{\psi}_k(\hat{\theta}_{k-1})$. It only requires storing $\{\varphi^{(j)}_k\}_j$ and $\{\zeta^{(j)}_{m,k}\}_{m,j}$  at each time step.
\smallskip

The next section provides a summary of the proposed algorithm.

\subsection{Algorithm Summary}\label{sec:algorithm_summary}
A summary of the estimation algorithm is collected in Algorithm~\ref{alg}. It starts from a given model parameterization as in \eqref{eq:model}, and considers the general case of estimating parameters in $G$, $f$, and the disturbance model. For clarity, it is given for the case of fixed sampling period $\Delta$ and constant input over the sampling interval. The algorithm is started with an initial value $\hat{\theta}_{0}\in \Theta$ that can be obtained by an a priori offline/patch identification, or using prior knowledge. The initial regressors $\varphi_0$ and $\{\varphi_0^j\}$ can be obtained from previous input-output data or simply set to zero.

Notice that, at each iteration, Lines \ref{alg:discretize_G} and \ref{alg:discretize_G_prime}  in  Algorithm~\ref{alg} require the discretization of the transfer functions $G(p;\hat{\theta}_{k-1})$ and $G'_j(p;\hat{\theta}_{k-1})$. Likewise, 
Lines \ref{alg:discretize_SDE} and \ref{alg:discretize_SDE_prime} require the discretization of the It\^{o}  stochastic differential equation of $w$ and its gradient with respect to $\theta$. This is done by evaluating appropriate exponential matrix functions of $\hat{\theta}_{k-1}$.  For irregular sampling times and general inputs, the discretization is to be done exactly by using the knowledge of the inter-sample behaviour of the input.

\IncMargin{1em}
\begin{algorithm}\small
\DontPrintSemicolon
\SetKwData{Left}{left}\SetKwData{This}{this}\SetKwData{Up}{up}
\SetKwInOut{Input}{input}
\SetKwInOut{Output}{output}
\SetKwProg{Init}{initializaion}{}{}
\SetKwFor{ForEach}{for each}{do}{end}
\Output{Sequence of estimates $\{\hat{\theta}_k\}_{k\geq 1}$}
\Input{Gain sequence $\{\gamma_k\}_{k\geq 1}$, positive integer $M\geq1$, initial parameter $\hat{\theta}_0$, sampling period $\Delta$, initial Hessian $R_0 = c I$ (for relatively large $c >0$),
       initial regressors $\varphi_0$ and $\{\varphi^{j}_0\}_{j=1}^{n_G}$, parameter vectors $\eta(\hat{\theta}_0)$ and $\{\eta_j(\hat{\theta}_0)\}_{j=1}^{n_G}$.}
\BlankLine
Set  $w_{m,0} = 0$ and $\zeta_{m,0}^{(j)} = 0$ for all $m$ and $j$\;
Set index $k\leftarrow1$ and collect data point $(y_1,u_1)$\;
\While{true}{
    $\beta_{m,k}^{(y)} \overset{\text{iid}}{\sim}\mathcal{N}(0,I_{n_w})$, $m =1, \dots, M$\;
    $w_{m,k} = A_\Delta(\hat{\theta}_{k-1}) w_{m,k-1} + B_\Delta(\hat{\theta}_{k-1}) \beta_{m,k}^{(y)}$\; \label{alg:discretize_SDE}
    $z_k  =  \varphi_{k-1}^\top \eta(\hat{\theta}_{k-1})$\; \label{alg:discretize_G}
    $\bar{y}_k = \frac{1}{M} \sum_{m=1}^M f\left(  z_k + C(\hat{\theta}_{k-1}) w_{m,k}\,; \hat{\theta}_{k-1}\right)$\;
    $z_k^{(j)} = [\varphi_{k-1}^{(j)}]^\top \eta_{j}(\hat{\theta}_{k-1})$\;\label{alg:discretize_G_prime}
    $\beta_{m,k}^{(\psi)} \overset{\text{iid}}{\sim}\mathcal{N}(0,I_{2n_w})$, $m =1, \dots, M$\;
    $\zeta_{m,k}^{(j)} = F^{(j)}_\Delta(\hat{\theta}_{k-1}) \zeta_{m,k-1}^{(j)} + L^{(j)}_\Delta(\hat{\theta}_{k-1}) \beta_{m,k}^{(\psi)}$\;\label{alg:discretize_SDE_prime}
    ${w}_{m,k} = [\zeta_{m,k}^{(j)}]_{1:n_w}$\;
    ${w}_{m,k}^{(j)} = [\zeta_{m,k}^{(j)}]_{n_w:2n_w}$\;
    $x_{m,k}^{(j)} = z_k^{(j)}+ C_j(\hat{\theta}_{k-1}) w_{m,k} + C(\hat{\theta}_{k-1}) w_{m,k}^{(j)}$\;
    $y^{(j)}_{m,k} = f'_j(x_{m,k}; \hat{\theta}_{k-1}) + f(x_{m,k};\hat{\theta}_{k-1})\, x_{m,k}^{(j)}$\;
    $\psi_{m,k} = -\begin{bmatrix} y^{(1)}_{m,k} & y^{(2)}_{m,k} & \dots & y^{(d)}_{m,k} \end{bmatrix}^\top$\;
    $\bar{\psi}_k = \frac{1}{M}  \sum_{m=1}^M \psi_{m,k}$\;
    $\varepsilon_k = y_k -\bar{y}_k$\;
    $R_k = R_{k-1} + \gamma_k \left[\bar{\psi}_k\bar{\psi}_k^\top - R_{k-1}\right]$\;
    $\hat{\theta}_k = \left[\hat{\theta}_{k-1} + \gamma_k R^{-1}_k \bar{\psi}_k\varepsilon_k\right]_\Theta$ \hfill\;
    store $\varphi_0$, $\{\varphi^{j}_0\}_{j=1}^{n_G}$ and  $\{w_{m,k}\}$,  $\{\zeta_{m,0}^{(j)}\}$  $\forall m, j$\;
    set index $k\leftarrow k+1$, and collect data $(y_k,u_k)$\;
    $\varphi_k = [ -z_{k} \,\, [\varphi_{k-1}]_{1:n-1}^\top \,\,  u_k \,\, [\varphi_{k-1}]_{n+2:2n-1}^\top]^\top$ \;
    $\varphi_k^{(j)} \!\!=\! [ -z_{k}^{(j)}  \,\, [\varphi^{(j)}_{k-1}]_{1:n-1}^\top \,\, u_k \,\, [\varphi^{(j)}_{k-1}]_{n+2:2n-1}^\top ]^\top$\!\;
} 
\caption{OE-QPEM online estimator}\label{alg}
\end{algorithm}\DecMargin{1em}

\subsection{Theoretical Motivation}\label{sec:theory}

The validity of the stochastic approximation step is achieved by the simulation of~\eqref{eq:model} using independent Wiener processes. Indeed, under this setting (and open-loop operation), \eqref{eq:unbiased_estimator-of_the_EF} is an unbiased estimator of  \eqref{eq:estimating_function}. 

The development in Section~\ref{sec:algorithm}, using \eqref{eq:asymptotic_cost}, implicitly assumes that $\{\varepsilon_k(\theta)\}$ is independent and weakly  stationary. This does not need to be the case: the validity of the stochastic approximation can be established for a more general class of (ergodic) statistically dependent prediction error processes; see \cite{Ljung1978sa}. This in particular means that the convergence of the algorithm can be established even for cases with under-modelling/misspecification, such that $\hat{\theta}_k \to \vartheta$ almost surely as $k\to\infty$ where $\vartheta$ is the set of roots of \eqref{eq:estimating_function}. Moreover, under mild regularity assumptions and an identifiability condition implying $\vartheta =\{\theta_\circ\}$, the asymptotic distribution $\sqrt{k}(\hat{\theta}_k - \theta_\circ)$ can be characterized; see \cite[App.11A]{Ljung1999}.

Additionally, interchanging the order of ordinary differentiation and stochastic integration can be justified (see \cite{hutton1984}), and therefore the gradient filters in \eqref{eq:grad_filters_w} and the gradients in \eqref{eq:grad_x} are well-defined.

For the sake of clarity in the previous sections, we used several additional assumptions that are stronger than what is actually required. The restriction to single-input single-output systems, black-box canonical parameterization of $G$, uniform sampling, and the assumption that the input is constant over the sampling interval are not needed. What is required is the knowledge of the inter-sampling behavior of the input. In addition,  the main requirement of the parameterization is its identifiability via the second-order moments of $y$. 

\section{Numerical Examples}\label{sec:examples}
In this section, we demonstrate the performance of the proposed algorithm using two numerical examples.
\subsection{Example 1}\label{ex1}
Consider the model
\begin{equation}\label{eq:model_ex1}
\begin{aligned}
    \diff x(t) &= a x(t) \diff t + b u(t) \diff t + \sigma \diff \beta(t),\\
    y(t)  &= x^2(t),
\end{aligned}
\end{equation}
where the measurement $y_k = y(t_k) + v_k$ is recorded with irregular sampling times. The sampling periods $\Delta_k = t_{k+1}-t_k$ are random with uniform distribution over the interval $[0.5, 1]$, and $v_k \sim \mathcal{N}(0,0.01^2)$. Let $\theta = \begin{bmatrix} a&b&\sigma\end{bmatrix}^\top$, and notice that the plant and disturbance models are jointly parameterized (they have the same pole). Consider a case where the data is generated by \eqref{eq:model_ex1} when the known input is a sum of 10 sinusoids: $u(t) = \sum_{\ell=1}^{10} A_\ell \cos(\omega_\ell t+\phi_\ell)$, with $A_\ell = 6$ for all $\ell$,  frequencies $\omega_\ell$ in $\{\frac{\pi}{5}, \frac{2\pi}{5}, \dots, {10\pi}\}$ selected uniformly at random, and Schroeder phases $\phi_\ell = \frac{\ell(\ell-1)}{10}\pi$. The true parameter $\theta_\circ = \begin{bmatrix} -1&1&1\end{bmatrix}^\top$, and the constraint set $\Theta :=\{\theta =[ a\; b\; \sigma]^\top \in \mathbb{R}^3 \;:\; a<0\}$; namely, only $a$ is constrained to guarantee stability. We applied Algorithm~\ref{alg} to ten independent data sets, using zero initial parameters, zero initial regressors $\varphi_0$, with $R_0 =5I$,  a gain sequence $\gamma_k = 1/k^{0.9}$ and constant $M =100$. 

\begin{figure}[ht]
    \centering
    \begin{subfigure}[b]{0.31\columnwidth}
        \includegraphics{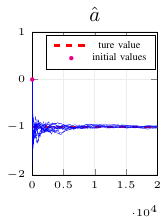}
    \end{subfigure}
    \,
    \begin{subfigure}[b]{0.31\columnwidth}        
        \includegraphics{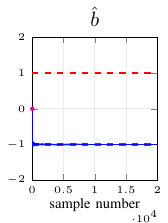}
    \end{subfigure}
    \,
    \begin{subfigure}[b]{0.31\columnwidth}
       \includegraphics{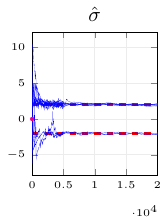}
    \end{subfigure}
     \caption{Ten MC simulations of Algorithm~\ref{alg} applied to \eqref{eq:model_ex1}.}
    \label{fig:ex1}
\end{figure}

The results are shown in Figure~\ref{fig:ex1} indicating the successful convergence of the algorithm. Notice that $b$ and $\sigma$ are identifiable only in magnitude due to the quadratic nonlinearity at the output.

\subsection{Example 2}
Consider now the model  given by
\begin{equation}\label{eq:model_ex2}
\begin{aligned}
\diff  w(t) &= \sigma\,  \diff  \beta(t)\\
x(t) &= \frac{c}{p^2  + a p + b } u(t) +  w(t)\\
y(t) &= \frac{1}{1 + |x(t)|^\alpha}
\end{aligned}
\end{equation}
and let $\theta = \begin{bmatrix} a& b& c& \sigma& \alpha \end{bmatrix}$.  The static nonlinear function in \eqref{eq:model_ex2} is known as the Hill function and is commonly used in biochemistry, particularly in pharmacology, to  describe the dose-response relationship \cite{Minto2000}.

Suppose the measurements $y_k = y(t_k) + v_k$ are recorded with a constant sampling period $\Delta = 0.5$, $v_k \overset{\text{iid}}{\sim} \mathcal{N}(0,0.05^2)$, and that the underlying data-generating system is given
with the true parameters $a_\circ= 1.2 $, $b_\circ = 0.27$, $c_\circ = 1 $, and the Hill coefficient $\alpha_\circ = 1.7$. The set $\Theta$ only constrains $a$ and $b$ such that the model is stable.

Consider first the following Gaussian disturbance in continuous-time
\begin{center}
\begin{tabular}{ c p{6.5cm} }
 Case 1: & $\diff  w(t) =  -0.75 w(t) \diff t + 1.5 \diff \beta(t)$
\end{tabular}
\end{center}
Notice that in this case, \eqref{eq:model_ex2} misspecifies the spectrum of the disturbance, while it matches the true parameterization of the plant model. 
We applied the proposed algorithm to fit \eqref{eq:model_ex2} to ten independent data sets, with random parameter initialization (uniformly within a 50\% interval of the true values), $R_0 = 10I$. The initial regressors were constructed using the first two data samples, and the algorithm started at the third sample. The gain sequence is $\gamma_k = 1/k^{0.85}$ and $M = 100$. In all cases, the input is pseudo-random binary input of amplitude $\pm5$ applied through a zero-order hold. 

\begin{figure}[ht]
    \centering
    \begin{subfigure}[b]{0.31\columnwidth}
        \includegraphics{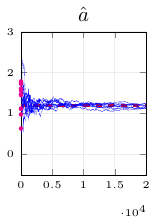}
    \end{subfigure}
    \begin{subfigure}[b]{0.31\columnwidth}
        \includegraphics{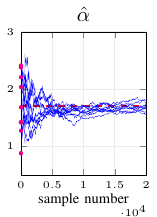}
    \end{subfigure}
    \begin{subfigure}[b]{0.31\columnwidth}
        \includegraphics{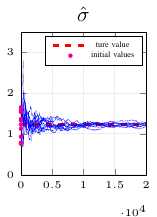}
    \end{subfigure}
    \caption{Ten  MC simulations of Algorithm~\ref{alg} applied to \eqref{eq:model_ex2} when the true disturbance model is given by Case 1.  The estimates of $b$ and $c$  (not shown) exhibit the same behaviour.}
    \label{fig:ex2a}
\end{figure}

The results in Figure~\ref{fig:ex2a} show that regardless of the specification of the disturbance model, the algorithm successfully converges to the true parameters. In particular, $\hat{\sigma}$ converges to the stationary marginal variance of $w$.

\begin{figure}[ht]
    \centering
    \makebox[0pt][c]{
    \begin{subfigure}[b]{0.48\columnwidth}
            \centering
            \includegraphics{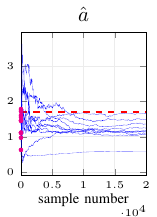}
    \end{subfigure}
    \begin{subfigure}[b]{0.48\columnwidth}
            \centering
            \includegraphics{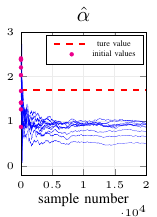}
    \end{subfigure}
    }
    \caption{Ten MC simulations of an online OE-QPEM algorithm that ignores  $w(t)$. The estimates of $b$ and $c$  (not shown) exhibit the same behaviour, and $\sigma$ is not estimated here.}
    \label{fig:ex2b}
\end{figure}

For comparison we also applied, to the same data set and settings, an online OE-QPEM algorithm ignoring $w(t)$ completely by assuming that $w(t) = 0$ for all $t$. The results are shown in Figure~\ref{fig:ex2b}, where the resulting bias is clear.

\begin{figure}[ht]
    \centering
    \begin{subfigure}[b]{0.31\columnwidth}
        \includegraphics{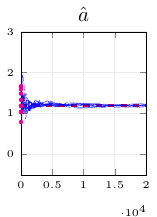}
    \end{subfigure}
    \begin{subfigure}[b]{0.31\columnwidth}
        \includegraphics{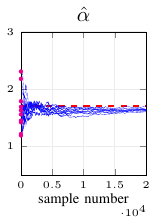} 
    \end{subfigure}
    \begin{subfigure}[b]{0.31\columnwidth}
        \includegraphics{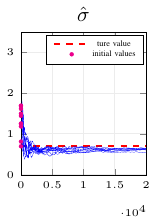}
    \end{subfigure}
    \caption{Ten  MC simulations of Algorithm~\ref{alg} applied to \eqref{eq:model_ex2} when the true disturbance model is given by Case 2. The estimates of $b$ and $c$  (not shown) exhibit the same behaviour.}
    \label{fig:ex2c}
\end{figure}

\begin{figure}[ht]
    \centering
    \begin{subfigure}[b]{0.31\columnwidth}
        \includegraphics{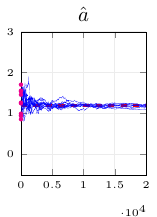}
    \end{subfigure}
    \begin{subfigure}[b]{0.31\columnwidth}
        \includegraphics{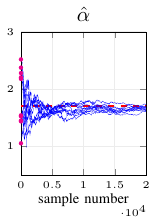}
    \end{subfigure}
    \begin{subfigure}[b]{0.31\columnwidth}
        \includegraphics{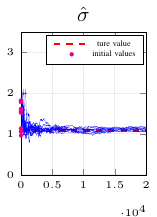}        
    \end{subfigure}
    \caption{Ten  MC simulations of Algorithm~\ref{alg} applied to \eqref{eq:model_ex2} when the true disturbance model is given by Case 3.  The estimates of $b$ and $c$ (not shown) exhibit the same behaviour.}
    \label{fig:ex2d}
\end{figure}

Finally,  to check the performance of the algorithm for cases under distributional misspecification we considered the following two additional cases for the disturbance model
 \begin{center}
\begin{tabular}{ c p{6.5cm} }
 Case 2: & $\diff  \xi(t) =  -0.75 \xi(t) \diff t + 1.5 \diff \beta(t)$ \newline $w(t_k) = \xi(t_k)\rho_k$,  and $\rho_k \sim \mathcal{U}(0,1)$ \\[0.75em]
 Case 3: &   $\diff  \xi(t) =  -0.75 \xi(t) \diff t + 1.5 \diff \beta(t)$\newline
 $w(t_k) =\!\!  \begin{cases}
     \xi(t_k) &\text{with prob. 0.8}\\
     w \sim \mathcal{N}(0, 0.5) &\text{with prob. 0.2}
 \end{cases} $
\end{tabular}
\end{center}
These cases correspond to mixed continuous-discrete non-Gaussian disturbances under which \eqref{eq:model_ex2} misspecifies both the marginal distribution and the dependence structure of $w(t)$. Distributional misspecification is not accounted for by the OE-QPEM estimator; hence, in these cases an asymptotic bias in its estimates is inevitable. The bias depends on the true nonlinearity, the input, and the moments of the true disturbance process. Still, as the results given in Figures~\ref{fig:ex2c} and \ref{fig:ex2d} show, the estimates converge to points close to the true values for the considered cases.

Applying an online OE-QPEM algorithm that ignores $w$ completely to these two cases yields a significant bias similar to that observed in Case 1 (see Figure~\ref{fig:ex2b}), and therefore the corresponding results are omitted.

\section{CONCLUSIONS}\label{sec:conclusions}
We propose a relatively simple online identification algorithm suitable for the identification of stochastic continuous-time parametric Wiener models from discrete sampled measurements. The proposed method is based on a stochastic approximation that approximates online, in a recursive fashion,  an output-error quadratic PEM estimator. The numerical simulation examples illustrate the convergence of the algorithm as expected, even when the disturbance model is incorrectly specified. We also show that misspecification in the dependence structure of the disturbance does not affect the convergence points of the plant and nonlinearity parameter estimates. A detailed asymptotic and finite-sample analysis of the proposed method is deferred to an extended future contribution.



\bibliographystyle{ieeetr}
\bibliography{library}

\begin{thebibliography}{10}

\bibitem{Hammond1965}
P.~Hammond, {\em Theory of self-adaptive control systems}.
\newblock Springer, 1965.

\bibitem{Eykhoff1974}
P.~Eykhoff, {\em System Identification. Parameter and State Estimation}.
\newblock John Wiley, 1974.

\bibitem{young2011}
P.~C. Young, {\em Recursive estimation and time-series analysis: An introduction for the student and practitioner}.
\newblock Springer Science \& Business Media, 2011.

\bibitem{Goodwin2014}
G.~Goodwin and K.~Sin, {\em Adaptive Filtering Prediction and Control}.
\newblock Dover Books on Electrical Engineering, Dover Publications, 2014.

\bibitem{chen2014}
H.-F. Chen and W.~Zhao, {\em Recursive identification and parameter estimation}.
\newblock CRC Press, 2014.

\bibitem{Ljung1983}
L.~Ljung and T.~S{\"o}derstr{\"o}m, {\em Theory and practice of recursive identification}.
\newblock MIT press, 1983.

\bibitem{Ljung1999}
L.~Ljung, {\em System Identification: Theory for the User}.
\newblock Prentice Hall, 2nd~ed., 1999.

\bibitem{Soederstroem1989}
T.~S{\"o}derstr{\"o}m and P.~Stoica, {\em System Identification}.
\newblock Prentice Hall, 1989.

\bibitem{Olsson2020}
J.~Olsson and J.~Westerborn~Alenl{\"o}v, ``Particle-based online estimation of tangent filters with application to parameter estimation in nonlinear state-space models,'' {\em Annals of the Institute of Statistical Mathematics}, vol.~72, pp.~545--576, 2020.

\bibitem{kantas2015}
N.~Kantas, A.~Doucet, S.~S. Singh, J.~Maciejowski, and N.~Chopin, ``On particle methods for parameter estimation in state-space models,'' {\em Statist. Sci.}, vol.~30, no.~3, pp.~328--351, 2015.

\bibitem{Singer2018}
H.~Singer, {\em Langevin and Kalman Importance Sampling for Nonlinear Continuous-Discrete State-Space Models}.
\newblock Springer, 2018.

\bibitem{Beskos2021}
A.~Beskos, D.~Crisan, A.~Jasra, N.~Kantas, and H.~Ruzayqat, ``Score-based parameter estimation for a class of continuous-time state space models,'' {\em SIAM Journal on Scientific Computing}, vol.~43, no.~4, pp.~A2555--A2580, 2021.

\bibitem{Surace2018}
S.~C. Surace and J.-P. Pfister, ``Online maximum-likelihood estimation of the parameters of partially observed diffusion processes,'' {\em IEEE {T}ransactions on {A}utomatic {C}ontrol}, vol.~64, no.~7, pp.~2814--2829, 2018.

\bibitem{Abdalmoaty2020b}
M.~Abdalmoaty, H.~Hjalmarsson, and B.~Wahlberg, ``The {G}aussian {M}aximum-{L}ikelihood {E}stimator {V}ersus the {O}ptimally {W}eighted {L}east-{S}quares {E}stimator [{L}ecture {N}otes],'' {\em IEEE Signal Processing Magazine}, vol.~37, no.~6, pp.~195--199, 2020.

\bibitem{oymak2019non}
S.~Oymak and N.~Ozay, ``Non-asymptotic identification of {LTI} systems from a single trajectory,'' in {\em 2019 American {C}ontrol {C}onference (ACC)}, pp.~5655--5661, IEEE, 2019.

\bibitem{simchowitz2020improper}
M.~Simchowitz, K.~Singh, and E.~Hazan, ``Improper learning for non-stochastic control,'' in {\em Conference on Learning Theory}, pp.~3320--3436, PMLR, 2020.

\bibitem{matni2019tutorial}
N.~Matni and S.~Tu, ``A tutorial on concentration bounds for system identification,'' in {\em 2019 IEEE 58th Conference on Decision and Control (CDC)}, pp.~3741--3749, IEEE, 2019.

\bibitem{li2023non}
Y.~Li, T.~Zhang, S.~Das, J.~Shamma, and N.~Li, ``Non-asymptotic system identification for linear systems with nonlinear policies,'' {\em arXiv preprint arXiv:2306.10369}, 2023.

\bibitem{foster2020learning}
D.~Foster, T.~Sarkar, and A.~Rakhlin, ``Learning nonlinear dynamical systems from a single trajectory,'' in {\em Learning for Dynamics and Control}, pp.~851--861, PMLR, 2020.

\bibitem{mania2020active}
H.~Mania, M.~I. Jordan, and B.~Recht, ``Active learning for nonlinear system identification with guarantees,'' {\em arXiv preprint arXiv:2006.10277}, 2020.

\bibitem{Bai2010book}
E.-W. Bai and F.~Giri, {\em Introduction to Block-oriented Nonlinear Systems}, pp.~3--11.
\newblock London: Springer London, 2010.

\bibitem{Boyd1985}
S.~Boyd and L.~Chua, ``Fading memory and the problem of approximating nonlinear operators with {V}olterra series,'' {\em IEEE Transactions on Circuits and Systems}, vol.~32, pp.~1150--1161, Nov 1985.

\bibitem{Wigren1993}
T.~Wigren, ``Recursive prediction error identification using the nonlinear {W}iener model,'' {\em Automatica}, vol.~29, no.~4, pp.~1011 -- 1025, 1993.

\bibitem{Wigren1994}
T.~Wigren, ``Convergence analysis of recursive identification algorithms based on the nonlinear {W}iener model,'' {\em IEEE Transactions on Automatic Control}, vol.~39, no.~11, pp.~2191--2206, 1994.

\bibitem{Wigren2023}
T.~Wigren, ``Recursive identification of a nonlinear state space model,'' {\em International Journal of Adaptive Control and Signal Processing}, vol.~37, no.~2, pp.~447--473, 2023.

\bibitem{Abdalmoaty2019}
M.~Abdalmoaty and H.~Hjalmarsson, ``Linear prediction error methods for stochastic nonlinear models,'' {\em Automatica}, vol.~105, pp.~49--63, 2019.

\bibitem{Abdalmoaty2018b}
M.~Abdalmoaty and H.~Hjalmarsson, ``Application of a linear {PEM} estimator to a stochastic {W}iener-{H}ammerstein benchmark problem,'' {\em IFAC-PapersOnLine}, vol.~51, no.~15, pp.~784 -- 789, 2018.

\bibitem{bereza2022stochastic}
R.~Bereza, O.~Eriksson, M.~R.-H. Abdalmoaty, D.~Broman, and H.~Hjalmarsson, ``Stochastic approximation for identification of non-linear differential-algebraic equations with process disturbances,'' in {\em 2022 IEEE 61st Conference on Decision and Control (CDC)}, pp.~6712--6717, IEEE, 2022.

\bibitem{Robbins1951}
H.~Robbins and S.~Monro, ``A stochastic approximation method,'' {\em Ann. Math. Statist.}, vol.~22, pp.~400--407, 09 1951.

\bibitem{Astroem1970}
K.~J. {\AA}str{\"o}m, {\em Introduction to Stochastic Control Theory}.
\newblock Mathematics in {S}cience and {E}ngineering, Academic Press, 1970.

\bibitem{Ljung1978sa}
L.~Ljung, ``Strong convergence of a stochastic approximation algorithm,'' {\em The Annals of Statistics}, vol.~6, no.~3, pp.~680--696, 1978.

\bibitem{hutton1984}
J.~E. Hutton and P.~I. Nelson, ``Interchanging the order of differentiation and stochastic integration,'' {\em Stochastic processes and their applications}, vol.~18, no.~2, pp.~371--377, 1984.

\bibitem{Minto2000}
C.~Minto, T.~Schnider, T.~Short, K.~Gregg, A.~Gentilini, and S.~Shafer, ``{Response Surface Model for Anesthetic Drug Interactions},'' {\em Anesthesiology}, vol.~92, pp.~1603--1616, 06 2000.

\end{thebibliography}

\end{document}